\documentclass{ws-ijtaf}
\usepackage{subfigure}
\usepackage{lineno}
\usepackage[usenames]{color}
\newcommand{\ud}{\mathrm{d}}

\newcommand{\ue}{\mathrm{e}}

\begin{document}

\markboth{G. Bormetti, V. Cazzola and D. Delpini}
{Option pricing under Ornstein-Uhlenbeck stochastic volatility: a linear model}

\catchline{}{}{}{}{}

\title{OPTION PRICING UNDER ORNSTEIN-UHLENBECK\\ STOCHASTIC VOLATILITY: A LINEAR MODEL}

\author{GIACOMO BORMETTI\footnote{Istituto Nazionale di Fisica Nucleare - Sezione di Pavia, via Bassi 6, Pavia, 27100, Italy}}

\address{
  Centro Studi Rischio e Sicurezza, Istituto Universitario di Studi Superiori\\
  V.le Lungo Ticino Sforza 56, Pavia, 27100, Italy\\
  Istituto Nazionale di Fisica Nucleare - Sezione di Pavia,\\
  via Bassi 6, Pavia, 27100, Italy\\
  \email{giacomo.bormetti@pv.infn.it}
}

\author{VALENTINA CAZZOLA}
\address{
  Centro Studi Rischio e Sicurezza, Istituto Universitario di Studi Superiori\\
  V.le Lungo Ticino Sforza 56, Pavia, 27100, Italy\\
  Dipartimento di Fisica Nucleare e Teorica, Universit\`a degli Studi di Pavia\\
  Istituto Nazionale di Fisica Nucleare - Sezione di Pavia\\
  via Bassi 6, Pavia, 27100, Italy\\
  \email{valentina.cazzola@pv.infn.it} 
}

\author{DANILO DELPINI}
\address{
  Dipartimento di Fisica Nucleare e Teorica, Universit\`a degli Studi di Pavia\\
  Istituto Nazionale di Fisica Nucleare - Sezione di Pavia\\
  via Bassi 6, Pavia, 27100, Italy\\
  \email{danilo.delpini@pv.infn.it} 
}

\maketitle

\begin{history}
\received{(Day Month Year)}
\revised{(Day Month Year)}
\end{history}

\begin{abstract}
We consider the problem of option pricing under stochastic volatility models,
focusing on the linear approximation of the two processes known as exponential Ornstein-Uhlenbeck and
Stein-Stein. Indeed, we show they admit the same limit dynamics in the regime of low
fluctuations of the volatility process, under which we derive the exact
expression of the characteristic function associated to the risk neutral
probability density. 
This expression allows us to compute option prices exploiting a formula derived by Lewis and Lipton. 
We analyze in detail the case of Plain Vanilla calls, being liquid instruments for which
reliable implied volatility surfaces are available. 
We also compute the analytical expressions of the first four cumulants, that are crucial to implement a 
simple two steps calibration procedure. It has been tested against a data set of options traded
on the Milan Stock Exchange. The data analysis that we present reveals a good fit with
the market implied surfaces and corroborates the accuracy of the linear approximation.
\end{abstract}


\section{Introduction}

The recent financial crisis has emphasized the need for reliable quantitative analysis 
of market data, able to guide the formulation of realistic theoretical models
for the dynamics of the traded assets. 
The Black-Scholes (B\&S) and Merton approach to option pricing~\cite{BS:1973,Merton:1973} 
assumes a Gaussian dynamics for the underlying assets and therefore it fails
to reproduce the well known stylized facts exhibiting clear evidence of
deviations from the normality assumption. This is why, in recent years,
more realistic alternative models have been proposed in the literature.
In particular, to capture the time varying nature of the volatility, assumed to be
constant in the B\&S approach, the stochastic volatility models (SVMs) represent
a theoretical framework used both in the research and the financial
practice. Among the most popular SVMs include the Heston~\cite{Heston:1993}, Stein-Stein (S2)~\cite{Stein_Stein:1991},
Sch\"oble-Zhu~\cite{SZ}, Hull-White~\cite{Hull:1987} and Scott~\cite{Scott:1987} models.
For reviews of SVMs we refer to~\cite{Micciche:2002,Lipton:2008,Mitra:2009}.
More recently, the model known in the econophysics literature as exponential
Ornstein-Uhlenbeck (ExpOU) has drawn particular attention because of its ability to reproduce a log-normal distribution
for the volatility, the so called leverage effect as well as the evidence of multiple time scales in the decaying of the volatility
auto-correlation function~\cite{MP:2006}. In~\cite{Borm}, the statistical
characterization of the process under the objective probability measure has been carried out
from both the analytical and numerical points of view. 

As far as the pricing problem is concerned, semi closed-form expressions for the price of European options are available
for the Heston and the S2 models.  For the ExpOU model the problem was addressed in the original paper by Scott,
who worked out a quasi closed-form pricing formula in the spirit of \cite{Hull:1987}. However, it eventually relies 
on the Monte Carlo (MC) simulation of the history of the volatility, while prices under Heston and S2 can be efficiently 
computed exploiting Fast Fourier Transform (FFT) numerical techniques~\cite{Carr:1999,Lewis:2001,Lipton:2001,Lord:2008}.
Recently, based on the Edgeworth expansion of the risk neutral density, an analytical expression of the pricing function for 
ExpOU has been derived~\cite{Perello:2007,PSM}. The accuracy of their approach has been tested numerically, and, at least for the considered regime, 
the approximate probability density function (PDF) they provide is unable to fit the one reconstructed via MC~\cite{Borm}. 
In conclusion, a satisfactory solution to the pricing problem under the ExpOU model is still lacking and 
it deserves further investigation. Indeed, the aim of the present paper is to discuss, under a risk neutral framework, its linear approximation  
for which a complete analytical characterization of the characteristic function (CF) can be provided. This allows to employ the Lewis and Lipton formula to 
efficiently solve for derivatives prices and to test the accuracy of the approximation in reproducing market observed volatility smiles. 

The paper is organized as follows. In Section~\ref{sec:themodel} we review the risk neutral formulation of
a SVM when the dynamics of the stochastic variable driving the volatility is described by the Ornstein-Uhlenbeck process, as for the ExpOU and the S2
models. We show how, under the regime of low fluctuations of the volatility process, 
the returns dynamics reduces to a linear one and we derive the exact analytical expression of the corresponding CF. As presented in \ref{AppA},
the explicit expressions of the first four cumulants, whose knowledge allows for an efficient calibration procedure, has been computed.
In Section~\ref{sec:NumRes} we perform a cross-sectional fitting of the Linear, ExpOU and S2 models, evaluating the
parameters from a data set of Plain Vanilla call options and detailing
the steps of the adopted calibration methodology. The ability of the Linear model to reproduce the data and the accuracy of the
approximation are evaluated comparing the volatility smiles reconstructed after calibration with the original market ones and with
those exhibited by the ExpOU and the S2 models. The final Section draws the relevant
conclusions and suggests some possible applications of the analytical results in the field of market risk management, such as Value-at-Risk
and Expected Shortfall evaluation.

\section{The Linear Model}\label{sec:themodel}

The class of SVMs we consider is described by the following system of stochastic differential equations (SDEs)
\begin{align}
  \label{eq:OUobj_dynamics}
    \ud S(t) &= \mu S(t) \ud t + \sigma(Y,t) S(t) \ud W_1(t)\,,~S(t_0) = S_0; \nonumber\\
    \ud Y(t) &= \alpha(\gamma - Y(t)) \ud t + k\rho \ud W_1(t)+ k \sqrt{1-\rho^2}
    ~\ud W_2(t)\,,~Y(t_0) = Y_0, 
\end{align}
where $\ud W_1$ and $\ud W_2$ are two independent Wiener processes,
while $S_0$, $\mu$, $Y_0$, $\alpha$, $\gamma$, $k$ and $\rho$ are
constant parameters. The dynamics of $Y(t)$ corresponds to an Ornstein-Uhlenbeck process, whose stationary mean and variance are given
by $\gamma$ and $\beta=k^2/(2\alpha)$. The rate of convergence to the steady state is given by $1/\alpha$, while
the correlation parameter $\rho$ takes value in $[-1,1]$.
The volatility $\sigma(Y,t)$ is a smooth function of $Y$ and $t$ and defining $\sigma(Y,t) = m \ue^{Y(t)}$ we obtain the ExpOU model \cite{Scott:1987,MP:2006}, while for
$\sigma(Y,t) = m Y(t)$ the S2 model~\cite{Stein_Stein:1991,SZ} is recovered. 

Given the market model specified by Eq.~\eqref{eq:OUobj_dynamics}, the standard approach to option
pricing consists of passing to an
equivalent risk neutral measure $\mathbb{P}^*$ under which
the discounted price process $\tilde{S}(t)\doteq \ue^{-r t} S(t)$, with 
$r$ the risk-free interest rate, is a martingale.
Indicating with $\mathbb{E}^*[\cdot]$ the expected value under
$\mathbb{P}^*$, this martingale property simply reads
$\mathbb{E}^*[\tilde{S}(t)|S(t_0)] = S(t_0)$ and the risk neutral dynamics of the model becomes
\begin{align}
    \ud S(t) &= r S(t) \ud t + \sigma(Y,t) S(t) \ud W_1^*(t),\nonumber\\
    \ud Y(t) &= \left[ \alpha (\gamma- Y(t))-k \eta(S,Y,t) \right] \ud t
    + k \rho \ud W_1^*(t) + k \sqrt{1-\rho^2} \ud W_2^*(t)
    \label{eq:rn_dY}.
\end{align}
In the above equation, $W_1^*(t)$ and $W_2^*(t)$ are independent standard
Brownian motions under the measure $\mathbb{P}^*$ and the  function $\eta(S,Y,t)$ correcting the
drift term of $Y(t)$ is called the \emph{market
price of volatility risk}, see \cite{FPS}. 
The function $\eta$ depends on the variables $t,S,Y$ and not on the
contract parameters. It takes the same form for different derivative contracts stipulated 
on the same underlying $S(t)$, parametrizes the space of risk neutral measures 
and defines $\alpha(\gamma-Y)-k\eta(S,Y,t)$ which is the \emph{risk neutral drift} of $Y$.

$S$ and $Y$ being Markovian processes, $\eta$ is a function of
the processes at time $t$, $\eta = \eta(t,S(t),Y(t))$; apart from suitable integrability conditions, from a
mathematical point of view $\eta$ is an arbitrary function and we assume it to be a
linear function of the process $Y(t)$
\begin{equation*}
    \eta(Y) = \eta_0 + \eta_1 Y(t).
\end{equation*}
In the light of its arbitrariness, our choice of a linear $\eta$ is eventually dictated
by the opportunity to preserve the mean reverting Ornstein-Uhlenbeck dynamics for $Y$ in the following, and 
not by any financial intuition.
It is worth noticing that this choice applies to the whole class of models~\eqref{eq:OUobj_dynamics},
independently on the explicit functional form of $\sigma(Y,t)$. 
It is coherent with the assumption done in \cite{PSM} and generalizes the one made by Stein and Stein in their original work~\cite{Stein_Stein:1991} 
corresponding to $\eta_1=0$.
We can redefine the parameters $\alpha$ and $\gamma$ as
\begin{equation}
    \alpha \rightarrow \tilde{\alpha} = \alpha + k \eta_1, \quad\mathrm{and}\quad 
    \gamma \rightarrow \tilde{\gamma} = \frac{\alpha \gamma - k \eta_0}{\tilde{\alpha}},\nonumber
\end{equation}
and, after applying It\^o's Lemma to the centred logarithmic return
$X(t)=\ln{S(t)} - \ln{S(t_0)}- r (t-t_0)$, finally the risk neutral dynamics reads
\begin{align}
    \label{eq:rndynamics}
    \ud X(t) & = -\frac{1}{2}\sigma^2(Y,t) \ud t + \sigma(Y,t) \, \ud W_1^*(t); \nonumber\\
    \ud Y(t) & = \tilde{\alpha} (\tilde{\gamma} - Y(t)) \ud t + k \rho \ud W_1^*(t)
    + k \sqrt{1-\rho^2} \ud W_2^*(t),
\end{align}
with initial conditions $X(t_0)=0$ and $Y(t_0)=Y_0$, and $\tilde{\alpha}>0$ that ensures the stationarity of the $Y$ process.

When the stationary variance of $Y$ is small, $\tilde{\beta} \doteq k^2 / (2 \tilde{\alpha})\ll 1$,
we can perform a first order Taylor expansion of $\sigma$ and $\sigma^2$ around $Y=\tilde{\gamma}$. Defining the process $Z=Y+1-\tilde{\gamma}$,
and the parameters $\tilde{m}=m\ue^{\tilde{\gamma}},~\tilde{k}=k$ for the ExpOU model, and $Z=Y/\tilde{\gamma}$,
$\tilde{m}=m\tilde{\gamma},~\tilde{k}=k/\tilde{\gamma}$ for the S2, the processes in~\eqref{eq:rndynamics} reduce to
\begin{align}
  \label{eq:Xlin_dynamics}
  \ud X(t) &=-\frac{\tilde{m}^2}{2}(2Z(t)-1)\ud t+\tilde{m} Z(t) \ud W_1^*~,~ X(t_0)=X_0=0, \\
  \label{eq:Zlin_dynamics}
  \ud Z(t) &=\tilde{\alpha} (1-Z(t))\ud t+\tilde{k}\rho\ud W_1^*(t) + \tilde{k}\sqrt{1-\rho^2}\ud W_2^*(t),Z(t_0)=Z_0.
\end{align}
The accuracy of the approximation has been discussed in detail for the ExpOU model in~\cite{Borm}.
In particular, a numerical analysis based on MC simulation for $\tilde{\beta} \lesssim 10\%$ supports the linearisation leading
to previous equations. However, the linear approximation does not preserve the martingality of $\tilde S(t)$, which 
is a crucial point for pricing purposes. From the definition $\tilde S(t)=S(t_0)\ue^{X(t)}$, the violation of this property is readily assessed by computing   
the deviation of $\mathbb{E}^*[\ue^{X(t)}|X_0]$ from 1. In order to obtain a martingale dynamics we modify the drift term driving $X(t)$
by means of a deterministic time dependent function $\mathcal{M}(t-t_0)$. The corrected SDE reads
\begin{align}
  \label{eq:martingale}
  \ud X(t) &=-\frac{\tilde{m}^2}{2}(2Z(t)-1+\mathcal{M}(t-t_0))\ud t+\tilde{m} Z(t) \ud W_1^*~.
\end{align}
The returns transition probability distribution $p_x(x,\tau|X_0,Z_0)$, with $\tau=t-t_0$, can be expressed in terms of the CF $f(\phi,\tau;X_0,Z_0)$, 
implicitly defined by
\begin{equation}\label{eq:pXcharacteristic}
  p_x(x,\tau|X_0,Z_0)=\frac{1}{2\pi}\int_{-\infty}^{+\infty}\ue^{-i \phi x}f(\phi,\tau;X_0,Z_0)\ud \phi ~ .
\end{equation}
The CF satisfies the Fokker-Planck backward equation in the Fourier space associated to the two-dimensional process described by 
Eqs.~\eqref{eq:Zlin_dynamics} and~\eqref{eq:martingale} .
This equation can be solved exactly, following a standard technique (see~\cite{Heston:1993,Masoliver_Perello:2002}), by guessing a solution of the
form
\begin{equation}\label{eq:cflinmodel}
   f(\phi,\tau; X_0, Z_0)= \exp \left\{-i\phi\frac{m^2}{2}\int_0^\tau \mathcal{M}(\tau')\ud\tau'+ A(\phi,\tau) +B(\phi,\tau)Z_0 + C(\phi,\tau)Z_0^2 + i\phi X_0 \right\} ~.
\end{equation}
The explicit expressions for the three functions $A$, $B$ and $C$ represent the main analytical result of the paper and read \footnote{
From now on we shall drop the tilde over the model parameters.}
\begin{align}
   A(\phi,\tau) &= \left[ \frac{h}{2} + 2\alpha \frac{n-h}{d} 
   + 2 k^2 \left( \frac{n-h}{d} \right)^2 + \frac{b-d}{4}\right]\tau \nonumber \\
   &-\frac{1}{2} \left[ \ln\left( 1 - g e^{-d \tau} \right) -
   \ln\left( 1 - g  \right) \right]\nonumber\\ 
   &- 2 k^2 \frac{e^{-d \tau} - 1}{\left( 1 - g \right)\left( 1 - g e^{-d \tau} \right)}
   \left\{ \frac{g}{d^3}\left[ \frac{\alpha}{2 k^2} (b+d) - h \right]^2 
   \right.\nonumber\\
   &\left. + \frac{( (g+1)h -2 n )^2 + 2(n-gh)(n-h)}{d^3} +
   \frac{g}{d^3}(n-h)^2\right\}\nonumber\\
   &-4 k^2\, \frac{ (g+1)h - 2 n }{d^3} \left( \frac{\alpha}{k^2} b - 2 h \right)
   \frac{\left( 1 + g e^{-\frac{d}{2}\tau} \right)\left( e^{-\frac{d}{2}\tau} - 1 \right)}
   {\left( 1 - g \right)\left( 1 - g e^{-d \tau} \right)}, \\
   B(\phi,\tau) &= 2~\frac{e^{-\frac{d}{2}\tau}\left[ (g+1)h-2n \right]
   + n + e^{-d \tau}(n-gh) - h} {d(1-g e^{-d \tau})},\\
   C(\phi,\tau) &= \frac{b-d}{4 k^2} \frac{1-e^{-d \tau}}{1-ge^{-d \tau}},
    \label{eq:Cpart}
\end{align}
where we have introduced the auxiliary functions $b\doteq 2\alpha(1 -i \rho\Phi)$,
$ d\doteq \sqrt{2\alpha^2\Phi^2+b^2}$, $ g\doteq(b-d)/(b+d)$, $h\doteq i\alpha m \Phi/k$, $n\doteq\alpha(b-d)/(2k^2)$,
and $\Phi\doteq km\phi/\alpha$.
It is relevant noting that the difference between principal logarithms in the second line has not been contracted into the
logarithm of the ratio. Indeed, this operation can be performed only by taking into account a suitable correction (see Eq.~(2.4) in~\cite{Veltman:1979}).
In order to assign the function $\mathcal{M}$, we impose $\mathbb{E}^*[\ue^{X(t)}|X_0]\equiv f(-i,\tau;X_0,Z_0)=1$ thus finding
\begin{align}
   \mathcal{M}(\tau) = \frac{2}{m^2}\frac{\ud}{\ud\tau}\left[A(-i,\tau) +B(-i,\tau)Z_0 + C(-i,\tau)Z_0^2 + X_0\right] ~.
\end{align}
Previous expression \emph{a posteriori} justifies the choice of $\mathcal{M}$ as an homogeneous function of time in Eq.~\eqref{eq:martingale}.

\section{Numerical Results}\label{sec:NumRes}

\subsection{Cross-sectional fitting}\label{sec:Calibration}

In the financial practice, SVMs are first calibrated on market data and then used for pricing. The calibration of the model parameters
could be performed following different approaches and this problem has been widely addressed in the literature. Several procedures have been
proposed in different contexts, see e.g. \cite{Lipton:2008,Mikhailov:2004,Galluccio:2005} and a discussion concerning asymptotic formulae
for the implied volatility smile can be found in~\cite{Forde:2009}, which also provides an overview of asymptotic methods.
In this work, we exploit the relationship between implied volatility smiles and 
the variance $\sigma_{\tau}^2$, skewness $\zeta_{\tau}$ and kurtosis $\kappa_{\tau}$ of the risk neutral PDF,
as provided by the expression given in \cite{Backus:2004} (see also \cite{Bouchaud:2000,Ciliberti:2008})
\begin{equation}\label{eq:BackusVola}
    \sigma_{\mathrm{imp},\tau}(d_1) \simeq \frac{\sigma_\tau}{\sqrt{\tau}} \left[ 1-\frac{\zeta_\tau}{3!}d_1-\frac{\kappa_\tau}{4!}(1-d_1^2)\right],
\end{equation}
where $\sigma_{\mathrm{imp},\tau}$ is the B\&S implied volatility for the
time to maturity $\tau$ and
\begin{equation}
d_1(\tau,K) \doteq \frac{\ln(S_0/K)+ r_\tau \tau+ \sigma_\tau^2 /2}{\sigma_\tau}\nonumber.
\end{equation}
The expression~\eqref{eq:BackusVola} is based on the approximation of the risk neutral PDF for fixed $\tau$ by means of a Gram Charlier expansion and
it does not rely on the choice of any specific underlying dynamics.
Its range of applicability is discussed in detail in
\cite{Backus:2004}, where it is shown that Eq.~\eqref{eq:BackusVola} is
effective for $d_1\sim 0$ and $\sigma_\tau \ll 1$, which is realized in practice (for $\tau \simeq 1$ year, typically $\sigma_{\tau}$ ranges from $.2$ to $.3$).

\begin{table}[t]\label{tab:MktData}
	\tbl{Implied volatilities market data.}
	{\begin{tabular}{@{}cccc@{}} \toprule
	    $\tau$ (yr)& $r_\tau$ ($\mathrm{yr}^{-1}$) & $\log(S_0/K)$ &
	    $\sigma_{\mathrm{imp},\tau}$ ($\mathrm{yr}^{-1/2}$)\\ \colrule
	
	0.0795	&	0.0425	&	\hphantom{-}0.0626	&	0.3354	\\
			&			&	\hphantom{-}0.0218	&	0.3089	\\
			&			&	-0.0175	&	0.2839	\\
			&			&	-0.0552	&	0.2599	\\
			&			&	-0.0657	&	0.2822	\\ \colrule
			
	0.1562	&	0.0465	&	\hphantom{-}0.1496	&	0.3427	\\
			&			&	\hphantom{-}0.0626	&	0.3114	\\
			&			&	\hphantom{-}0.0218	&	0.2823	\\
			&			&	-0.0175	&	0.2700	\\
			&			&	-0.0552	&	0.2566	\\
			&			&	-0.0916	&	0.2592	\\
			&			&	-0.1267	&	0.2630	\\
			&			&	-0.1606	&	0.2686	\\ \colrule
			
	0.2329	&	0.0474	&	\hphantom{-}0.0626	&	0.3347	\\
			&			&	\hphantom{-}0.0218	&	0.2874	\\
			&			&	-0.0175	&	0.2704	\\
			&			&	-0.0552	&	0.2726	\\
			&			&	-0.0916	&	0.2681	\\
			&			&	-0.1267	&	0.2593	\\
			&			&	-0.1606	&	0.2643	\\ \colrule

			
	0.3260		&		0.0471	&	0.1496	&	0.4210          	\\
			&			&	0.0626	&	0.4626	\\
			&			&	\hphantom{-}0.0218	&	0.2729	\\
			&			&	-0.0175	&	0.2718	\\
			&			&	-0.0552	&	0.2669	\\
			&			&	-0.0916	&	0.2616	\\
			&			&	-0.1267	&	0.2603	\\
			&			&	-0.1606	&	0.2578	\\ \colrule
			
	0.5781		&	0.0469	&	\hphantom{-}0.0218	&	0.2992	\\
			&			&	-0.0175	&	0.2949	\\
			&			&	-0.0552	&	0.2898	\\
			&			&	-0.0916	&	0.2817	\\
			&			&	-0.1267	&	0.2801	\\
			&			&	-0.1606	&	0.2799	\\ \colrule
			
	0.8274	&	0.0468	&	-0.0552	&	0.2966	\\
			&			&	-0.0916	&	0.2919	\\
			&			&	-0.1267	&	0.2865	\\
			&			&	-0.1606	&	0.2823	\\  \botrule
	\end{tabular}}
\end{table}
We can compute the cumulants for the Linear model exploiting the analytical formulae provided by Eq.~\eqref{eq:cflinmodel}-\eqref{eq:Cpart}, to
which they are related through
\begin{equation}
  k_{n,\tau}=\left( -i \right)^n \frac{\partial^n \ln f (\phi,\tau; X_0, Z_0)}{\partial \phi^n}\Bigg|_{\phi=0} ~ .
\end{equation}
To obtain the analytical expressions, using \texttt{MATHEMATICA}\textsuperscript{\textregistered} we approximate 
the logarithm of $f$ by means of a 4-th order Taylor expansion around $\phi=0$ and then we extract the four coefficients
of the expansion and multiply them by the appropriate constant factor, finally finding the results reported in \ref{AppA}.
After identifying $\sigma^2_{\tau}$ with $k_{2,\tau}$, then the skewness and kurtosis read
$\zeta_\tau = k_{3,\tau} / \sigma_{\tau}^{3}$, and $\kappa_\tau = k_{4,\tau} / \sigma_{\tau}^4$, respectively.
The exact analytical CF is also available for the S2 model (see~\cite{Stein_Stein:1991,SZ}) and following the same approach
it is possible to compute explicitly the related expressions for the cumulants.

We limit our analysis to Plain Vanilla call options, whose implied volatilities
$\sigma_{\mathrm{imp},\tau}$ are available from market data providers for different
maturities $\tau$ and strike prices $K$. The underlying spot price $S_0$ and the term structure of risk-free rates $r_\tau$
can be retrieved from the market as well.
Table~1 sums up the complete data set available, corresponding to options written on the Intesa San Paolo S.p.A. asset with spot 
price $S_0 = 5.16$ EUR, as of 22nd November 2007 on the Milan Stock Exchange. Annualized implied volatilities values are quoted, with the 
corresponding log-moneyness, time to maturities and risk-free rates retrieved from the EUR yield curve.
The complete list of parameters to calibrate is given by $m$, $Y_0$ (equivalently $Z_0$), $k$, $\alpha$, $\gamma$, and $\rho$, see Eq.~(\ref{eq:rndynamics}).
Under the hypothesis that the process $Y(t)$ driving the volatility has reached the stationary state, we fix $Y_0 = \gamma$ with $\gamma = 0$ for
the Linear and ExpOU models, and $\gamma = 1$ for S2, see also~\cite{Borm}. In order to fit the remaining four free parameters, we adopt the calibration
procedure detailed below.

\begin{figure}[h!]
   \begin{center}
      {\includegraphics[width=0.75\textwidth]{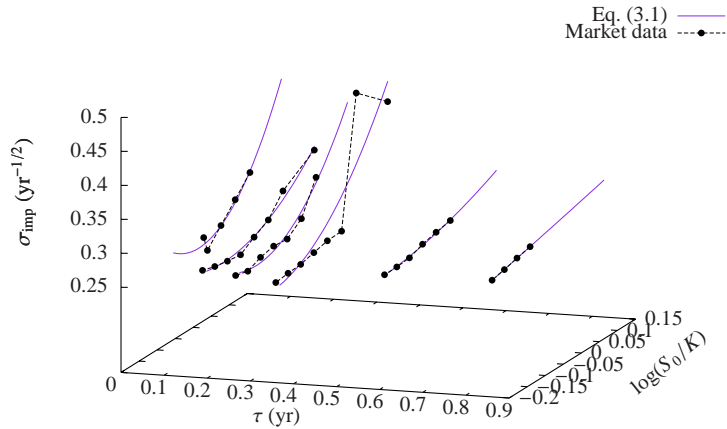}}
   \end{center}
   \caption{Implied volatilities for Intesa San Paolo: market data (dark points) and parabolic approximation, see Eq.~(\ref{eq:BackusVola}).}
   \label{fig:Backus_surface}
\end{figure}

{\it Fit $\sigma_\tau$, $\zeta_\tau$ and $\kappa_\tau$ from market smiles.}
Exploiting Eq.~(\ref{eq:BackusVola}), which provides an approximation
of the smiles in a suitable region around $d=0$, we fit the empirical data with a Marquard-Levenberg algorithm~\cite{Press:1989}
retrieving the optimal values $\sigma_\tau^{Mk}$, $\zeta_\tau^{Mk}$, and $\kappa_\tau^{Mk}$.
These values and the associated standard errors $\epsilon_{\sigma_\tau}^{Mk}$, $\epsilon_{\zeta_\tau}^{Mk}$, and $\epsilon_{\kappa_\tau}^{Mk}$, are
summarized in Table~2. The last column shows that smiles made of fewer points, like the one corresponding to $\tau=0.8274$ (4 points), result
in a greater estimation error, but we expect this effect to be greatly reduced for larger data sets, if available.
\begin{table}[h]\label{tab:MLAlgo}
	\tbl{Market calibrated normalized cumulants and their standard errors.}
	{\begin{tabular}{@{}c|cc|cc|cc@{}} \toprule
	\multicolumn{1}{c}{$\tau$ (yr)}	&\multicolumn{2}{c}{$\sigma_\tau^{Mk}\pm\epsilon_{\sigma_\tau}^{Mk}$}&\multicolumn{2}{c}{$\zeta_\tau^{Mk}\pm\epsilon_{\zeta_\tau}^{Mk}$}&\multicolumn{2}{c}{$\kappa_\tau^{Mk}\pm\epsilon_{\kappa_\tau}^{Mk}$}	\\	\colrule
	
	0.0795&	0.0885			&	0.0063			&	-0.80\hphantom{0}&	0.20\hphantom{0}&	2.0\hphantom{0}&	2.1\hphantom{0}\\	
	0.1562&	0.1145			&	0.0012			&	-0.578			&	0.064			&	1.44			&	0.31			\\	
	0.2329&	0.164\hphantom{0}&	0.013\hphantom{0}&	-1.11\hphantom{0}&	0.16\hphantom{0}&	4.6\hphantom{0}&	1.8\hphantom{0}\\	
	0.3260&	0.210\hphantom{0}&	0.071\hphantom{0}&	-1.82\hphantom{0}&	0.92\hphantom{0}&	5.3\hphantom{0}&	7.8\hphantom{0}\\	
	0.5781&	0.235\hphantom{0}&	0.011\hphantom{0}&	-0.587			&	0.066			&	1.7\hphantom{0}&	1.2\hphantom{0}\\	
	0.8274&	0.269\hphantom{0}&	0.011\hphantom{0}&	-0.760			&	0.068			&	0.2\hphantom{0}&	1.0\hphantom{0}\\ \botrule
	
	\end{tabular}}
\end{table}
In Fig.~\ref{fig:Backus_surface} we present the market data and the parabolic approximation Eq.~(\ref{eq:BackusVola}) with parameters fixed as in
Table~2. Individual smiles are very well reproduced and for long time to maturities the curves flatten, as expected, while
the highest implied kurtosis corresponds to the shortest $\tau$. For $\tau=0.3260$ yr we notice that the volatilities for
extreme positive log-moneyness are suspiciously out of scale. We calibrate on the entire data set, but we expect that this large
fluctuations will not be reproduced by the models under investigation. 

{\it Find optimal $\alpha$, $k$, $m$, and $\rho$ from the time scaling of $\sigma_\tau$, $\zeta_\tau$ and $\kappa_\tau$.} 
The calibration can be done by fitting the values reported in Table~2 with those computed from the models.
The scaling of $\sigma_\tau$, $\zeta_\tau$ and $\kappa_\tau$ with $\tau$ is known analytically for the Linear and S2 models, while it can be estimated numerically for ExpOU.
For the latter case, we sample $N_{MC}=10^{5}$ paths and we obtain the MC estimators $\sigma_\tau^{MC}$, $\zeta_\tau^{MC}$,
$\kappa_\tau^{MC}$ with associated standard errors $\epsilon_{\sigma_\tau}^{MC}$, $\epsilon_{\zeta_\tau}^{MC}$ and $\epsilon_{\kappa_\tau}^{MC}$.
The set of optimal values satisfies the equation 
\begin{equation*}
   \alpha^*,k^*,m^*,\rho^* =\mathop{\mathrm{argmin}}_{\alpha,k,m>0,~\rho\in[-1,1]}\sum_{\tau}
  \left[
  \frac{(\sigma_\tau^{Mk}-\sigma_\tau)^2}{{\epsilon^2_{\sigma_\tau}}}
  +\frac{(\zeta_\tau^{Mk}-\zeta_\tau)^2}{{\epsilon^2_{\zeta_\tau}}}
  +\frac{(\kappa_\tau^{Mk}-\kappa_\tau)^2}{{\epsilon^2_{\kappa_\tau}}}
  \right] ~ ,
\end{equation*}
where $\epsilon^2_{\sigma_{\tau}} = {\epsilon_{\sigma_\tau}^{Mk}}^2 + {\epsilon_{\sigma_\tau}^{MC}}^2$,
and analogously for $\epsilon^2_{\zeta_{\tau}}$ and $\epsilon^2_{\kappa_{\tau}}$, the MC error
being zero for the Linear and S2 models since their cumulants are known analytically and no MC simulation is required.
The optimization problem is solved by means of \verb+MINUIT+ routines \cite{Minuit} and the 
final results are contained in Table~3. The value of $\beta$ of order $10\%$ from the calibration of the ExpOU dynamics supports the
accuracy of the linear approximation, while no statistically significant differences are observed between the Linear and the S2 models.
\begin{table}[h]
	\tbl{Optimal parameters for the three models.}
	{\begin{tabular}{@{}c|cc|cc|cc|cc|cc@{}} \toprule
	\multicolumn{1}{c}{} &\multicolumn{2}{c}{$\alpha^*\pm\epsilon_{\alpha}$}&\multicolumn{2}{c}{$k^*\pm\epsilon_{k}$ }&\multicolumn{2}{c}{$m^*\pm\epsilon_{m}$}&\multicolumn{2}{c}{$\rho^*\pm\epsilon_{\rho}$}&\multicolumn{2}{c}{$\beta^*\pm\epsilon_{\beta}$}	\\
	\multicolumn{1}{c}{} &\multicolumn{2}{c}{($\mathrm{yr}^{-1}$)}&\multicolumn{2}{c}{($\mathrm{yr}^{-1/2}$)}&\multicolumn{2}{c}{($\mathrm{yr}^{-1/2}$)}&\multicolumn{2}{c}{}&	\multicolumn{1}{c}{}	\\	\colrule
	
	ExpOU	&	6.3	& 1.5		&	1.3	& 0.1		&	0.266	& 0.018		&	-0.51 & 0.09	&
	0.13 & 0.04 	\\
	S2		&	5.7	&	1.3	&	1.9	&	0.4	&	0.265	&	0.008	&	-0.41            &	0.07  &
	0.32 & 0.14 	\\
	Lin		&	5.6	&	1.3	&	1.9	&	0.4	&	0.264	&	0.008	&	-0.41			&	0.07    &
	0.34 & 0.15 	\\ \botrule
	
	\end{tabular}}
	 \label{tab:3ModParam}
\end{table}

As a final comment, we point out that the most computationally instensive step of the above procedure corresponds to the MC simulation of
the ExpOU dynamics, which was performed on a i686 machine equipped with Intel(R) Core(TM)2 Quad CPU Q6600  @ 2.40GHz
processor, taking about 1500 minutes to complete. Taking advantage of
the analytical cumulants, the calibration of the other two models does not suffer this limitation and requires only a few seconds.

\subsection{Option prices and implied volatility smiles}\label{s:numprice}

With the parameters values reported in Table~3 we compute the option prices, extract the implied volatilities and plot
the reconstructed volatility smiles against the market ones in order to asses the accuracy of the Linear model. We also compare
the results with the smiles generated by the ExpOU and S2 models.

The option prices for the ExpOU model have to be MC computed, while in the other cases the knowledge of the CF allows to
use the pricing formula~\cite{Lewis:2001,Lipton:2001}
\begin{eqnarray}\label{eq:lewisprice}
  C(S_0,t_0)&=&-\frac{S_0}{2\pi}\ue^{-D}\int_{i c - \infty}^{i c + \infty} \ud z ~ \ue^{-i z D} ~
  \frac{f(-z)}{z^2 - i z}\nonumber\\
  &=&-\frac{S_t}{2\pi}\ue^{D(c-1)}\left\{\int_{0}^{+ \infty} \ud \omega ~\cos(\omega D)~
  \mathrm{Re}\left[W(\omega_+)f(-\omega_+)+W(\omega_-)f(-\omega_-)\right]
  \right.\nonumber\\
  &&\qquad\qquad\quad\left.+\int_{0}^{+ \infty} \ud \omega~\sin(\omega D)~
  \mathrm{Im}\left[W(\omega_+)f(-\omega_+)-W(\omega_-)f(-\omega_-)\right]
  \right\}~,\nonumber\\
\end{eqnarray}
where we have dropped the dependence on $\tau$, $X_0$ and $Z_0$ in the characteristic function $f$. In the previous formula,
$D=\log{(S_0/K)}+r\tau$, $S_0$ is the spot price, $K$ the strike, $W(\omega)=[\omega^2-i\omega]^{-1}$, $\omega_+=+\omega+i c$,
$\omega_-=-\omega+i c$, and $z=\omega + i c$. When $\rho \in (-1,1)$, for the Linear model Eq.\eqref{eq:lewisprice} applies for
$c\in \left\{ c: c > 1 \right\} \cap \left\{ c : c_- < c < c_+ \right\}$, where $c_{\mp}= \alpha / [k m (\rho \mp 1)]$ correspond
to the imaginary part of the singularities of the CF~\eqref{eq:cflinmodel}.
For $\rho\sim -0.5$, $\alpha\sim 6$, $k\sim 2$, and $m\sim 0.26$ the above intersection is not empty and
we set $c=\lambda \frac{\alpha}{k m}\frac{1}{1+\rho}$ with $\lambda=0.5$ (we have verified that our results are insensitive to different
choices of $\lambda\in(\frac{km}{\alpha}(1+\rho),1)$). The identification of the singularities and a similar analysis has been performed
for the characteristic function of the S2 model.

We compute the integrals involved in Eq.~\eqref{eq:lewisprice} using an adaptive trapezoidal algorithm,
optimized to calculate sine or cosine transforms, e.g. see the routine  \verb+dqawf.f+ available at \emph{http://www.netlib.org/quadpack/}.
In order to check the reliability of the above numerical setup we have compared the implied volatilities obtained through
inversion of the prices computed via~\eqref{eq:lewisprice} with those from a MC
simulation ($N_{MC}=10^7$, parameters fixed as in Table 3) of the Linear model.
For all the available time to maturities, the implied volatility smiles and the simulated ones are in full agreement at 68\% confidence level.\par

\begin{figure}[h!]
   \begin{center}
      \subfigure
      {\includegraphics[width=0.45\textwidth]{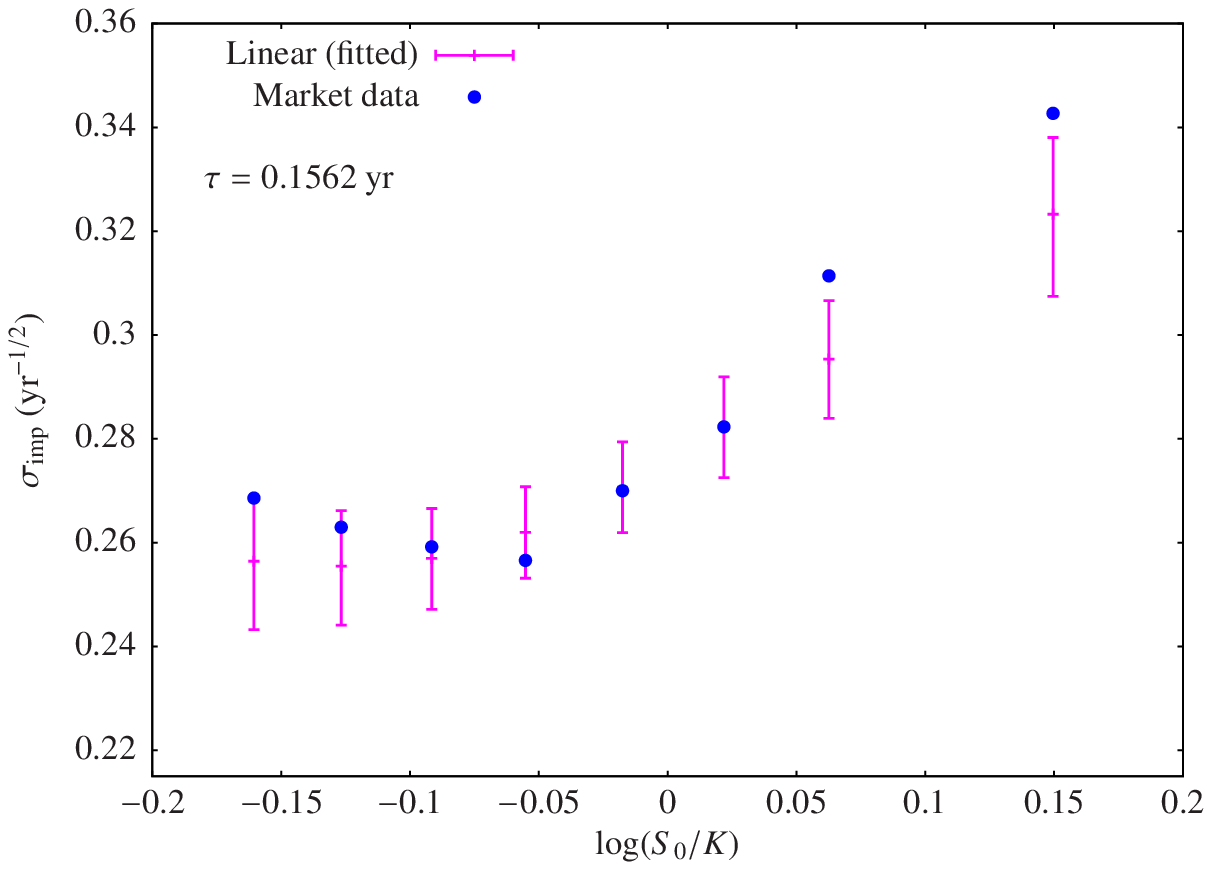}}
      \hspace{5mm}
      {\includegraphics[width=0.45\textwidth]{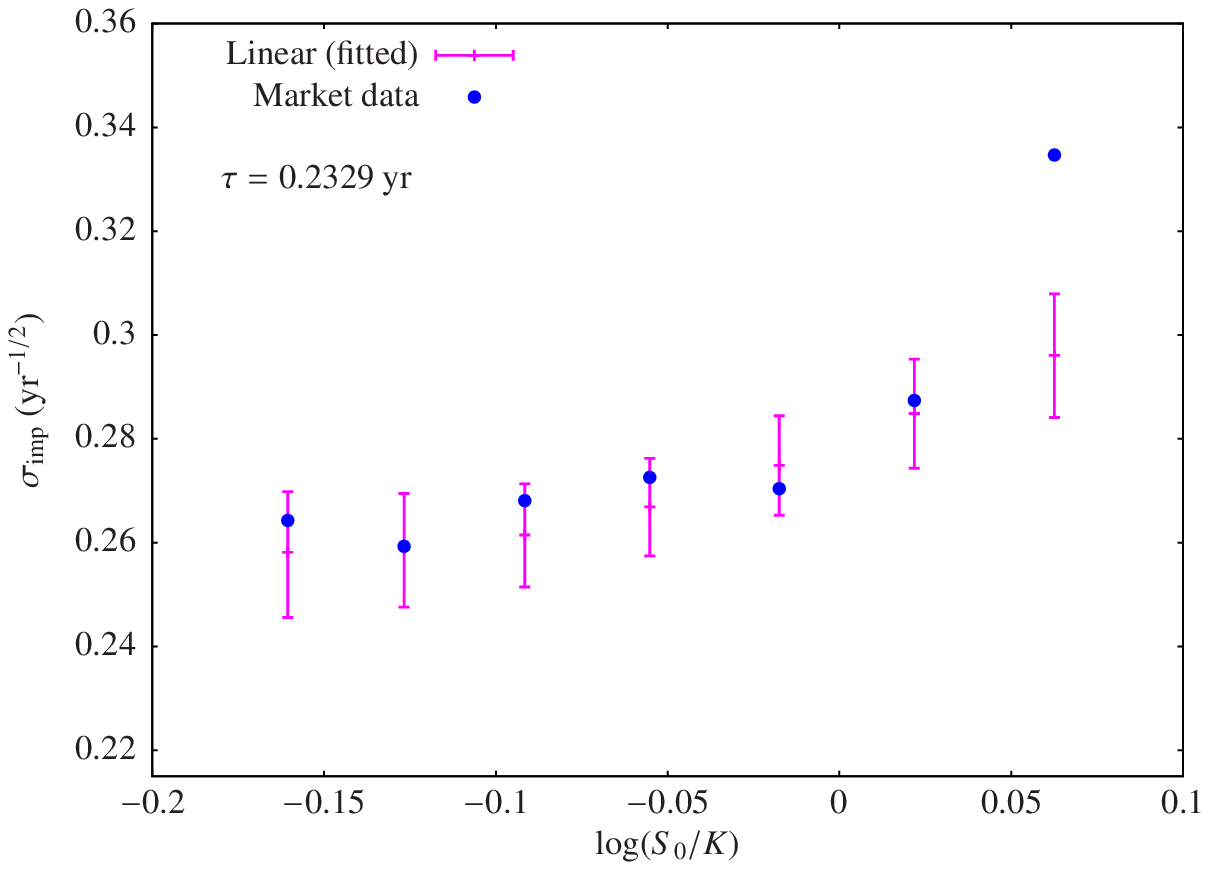}}
      \\
      {\includegraphics[width=0.45\textwidth]{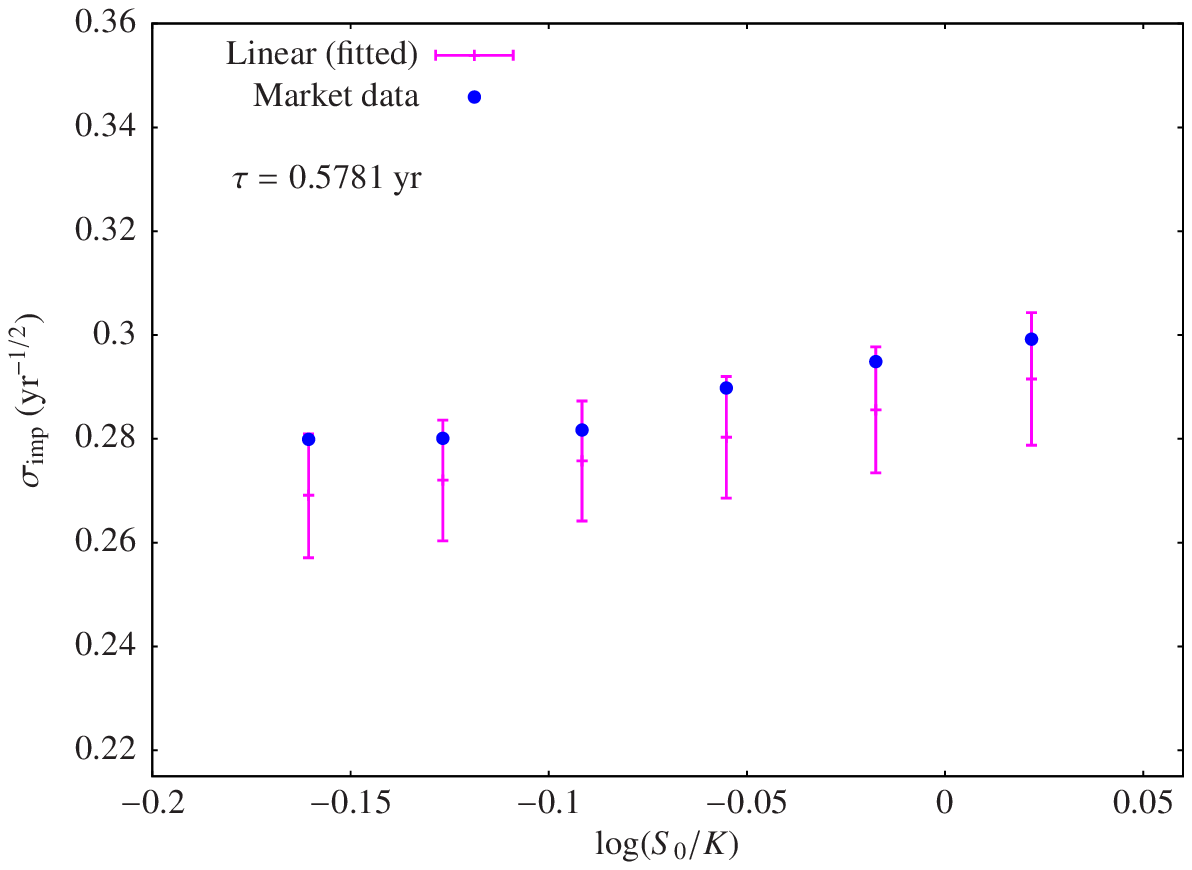}}
      \hspace{5mm}
      {\includegraphics[width=0.45\textwidth]{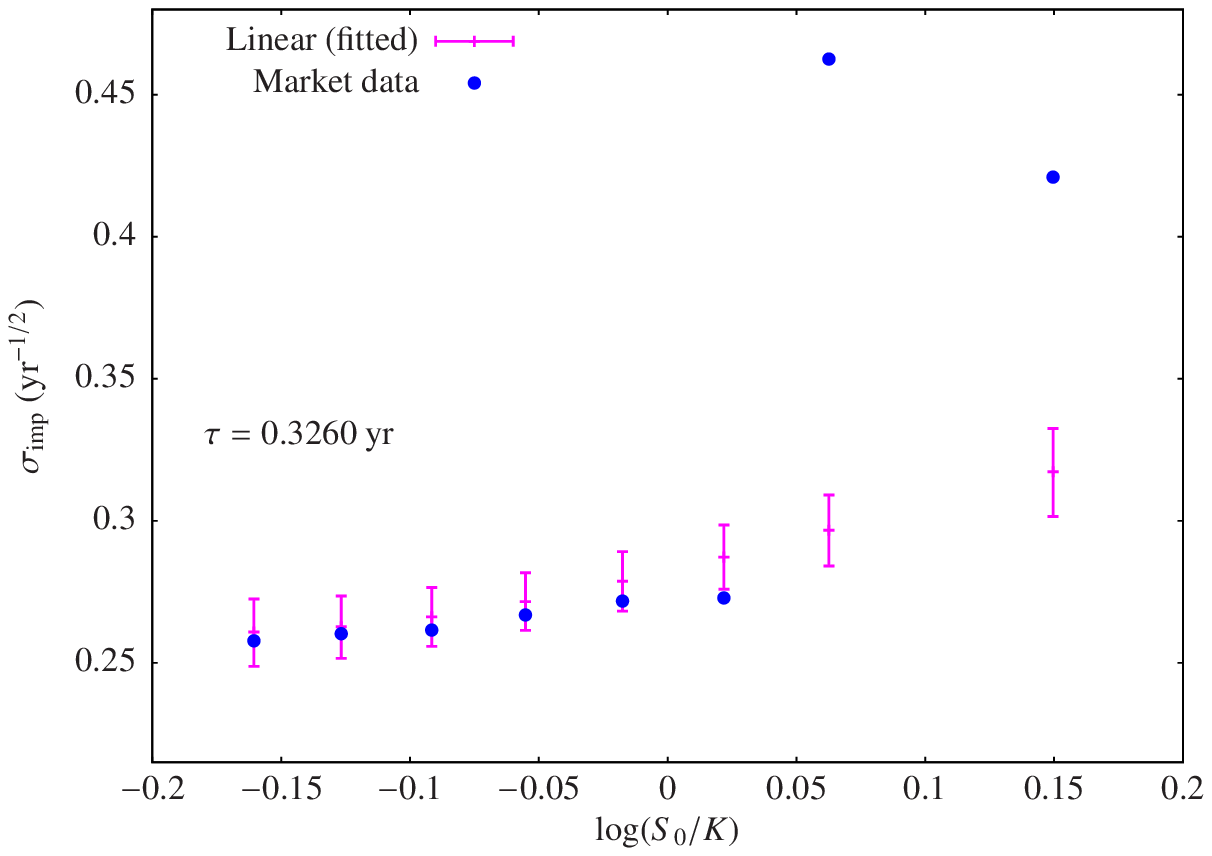}}
   \end{center}
   \caption{Comparison between market implied volatility smiles for Intesa San Paolo
   and those obtained from the Linear model. Parameters values as in the third row of Table 3, as obtained from
   the calibration of the Linear model.}
   \label{fig:Lsmiles}
\end{figure}

In Fig.~\ref{fig:Lsmiles} we present the implied volatilities smiles reconstructed by the Linear model, showing
its ability to capture the correct shape of the volatility. The agreement slightly decreases for deep in and out of the money
options; from the third panel corresponding to $\tau = 0.326~\mathrm{yr}$ we notice that the fluctuations of the two outlying points are not reproduced,
as expected.
The statistical uncertainty on the parameters values reflects in the standard errors
associated to the curves. The error propagation has been performed making use of the first order derivatives computed by means of finite difference
methods.
It is worth pointing out that, following the guidelines depicted in the previous Section,
calibration and pricing are carried out in an efficient way; this is mainly due to the analytical characterization we provided for the CF and cumulants,
making the entire approach a real time procedure.
\begin{figure}[h!]
   \begin{center}
      \subfigure
      {\includegraphics[width=0.45\textwidth]{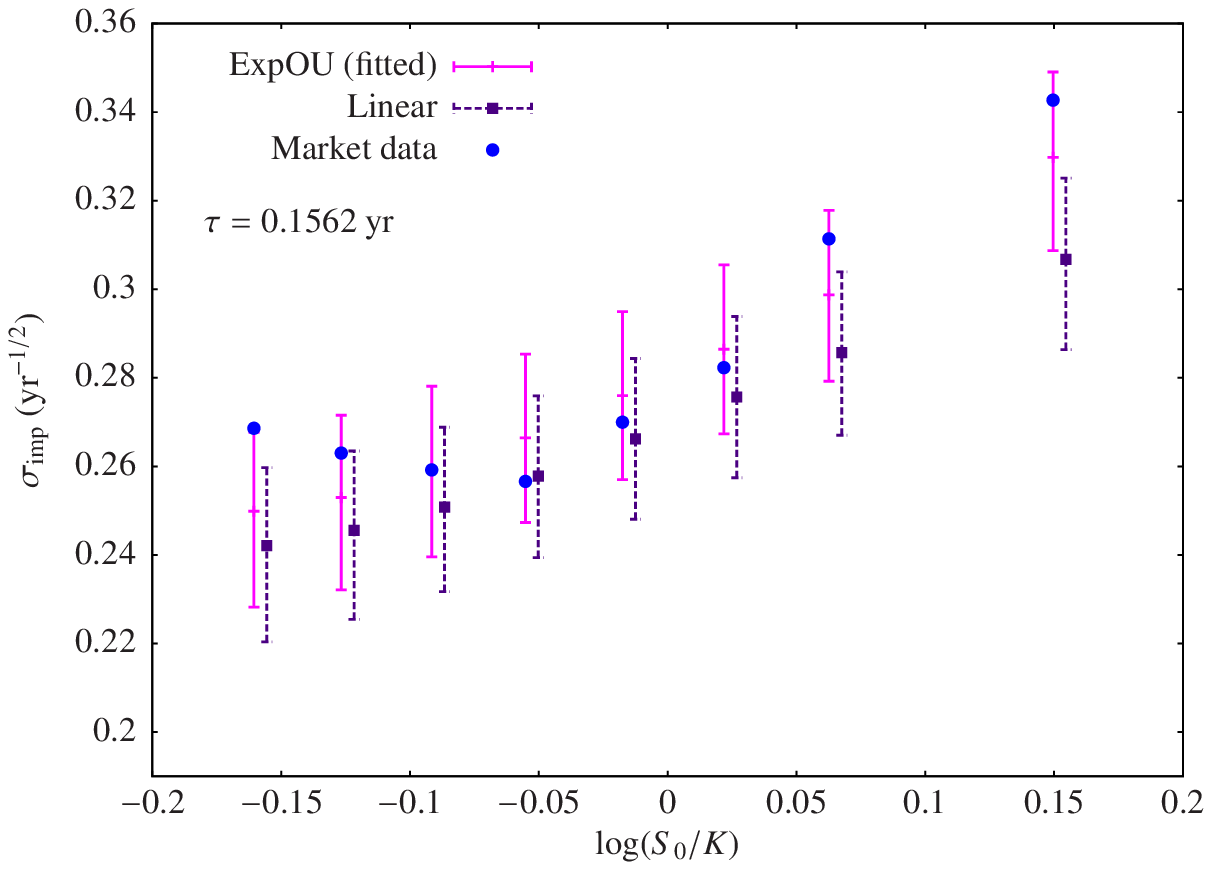}}
      \hspace{5mm}
      {\includegraphics[width=0.45\textwidth]{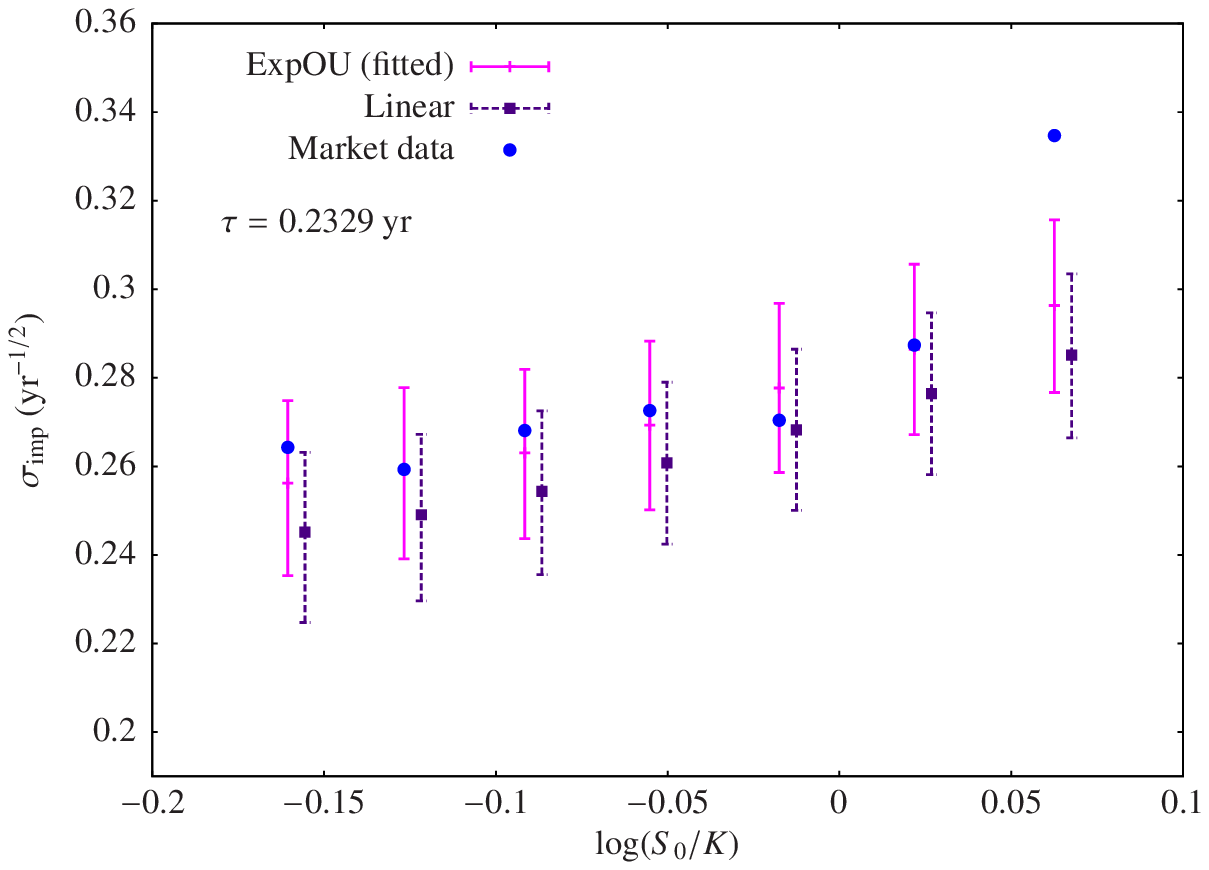}}
      \\
      {\includegraphics[width=0.45\textwidth]{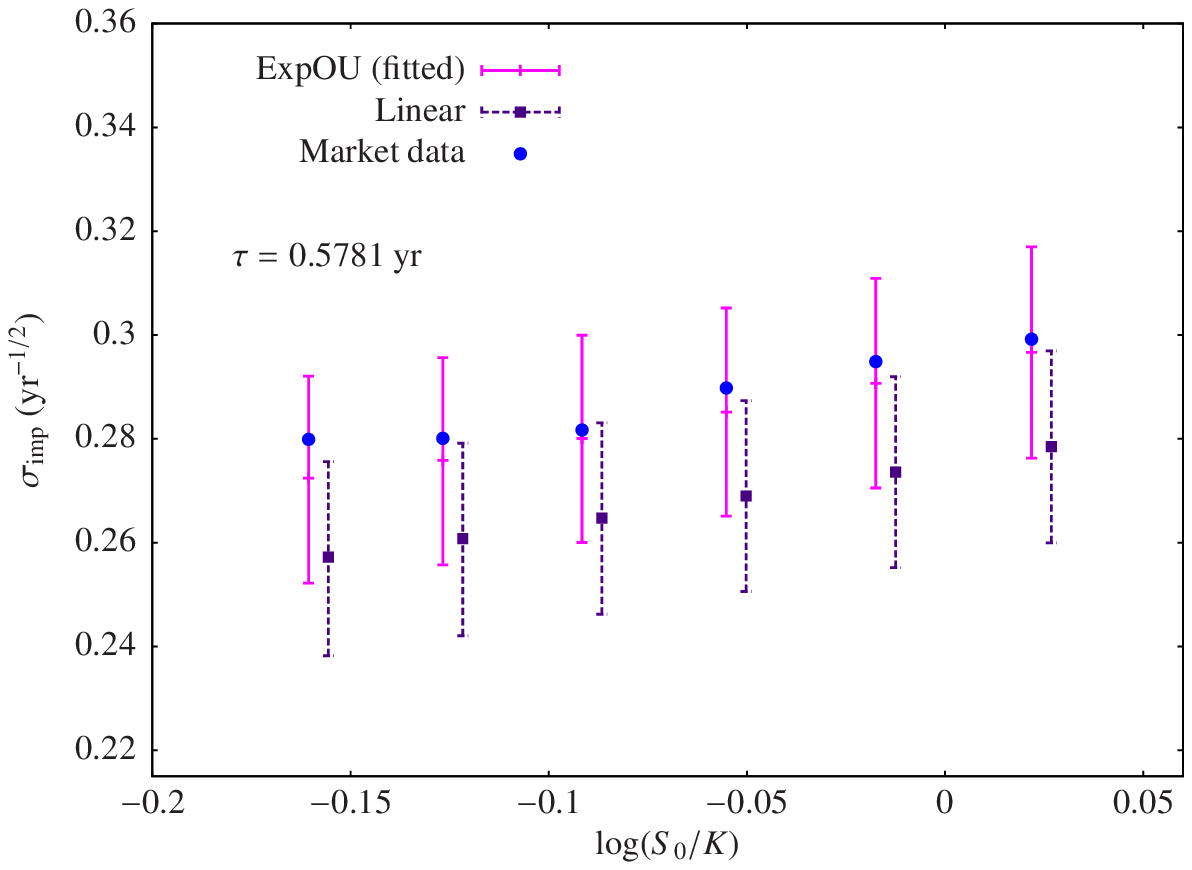}}
      \hspace{5mm}
      {\includegraphics[width=0.45\textwidth]{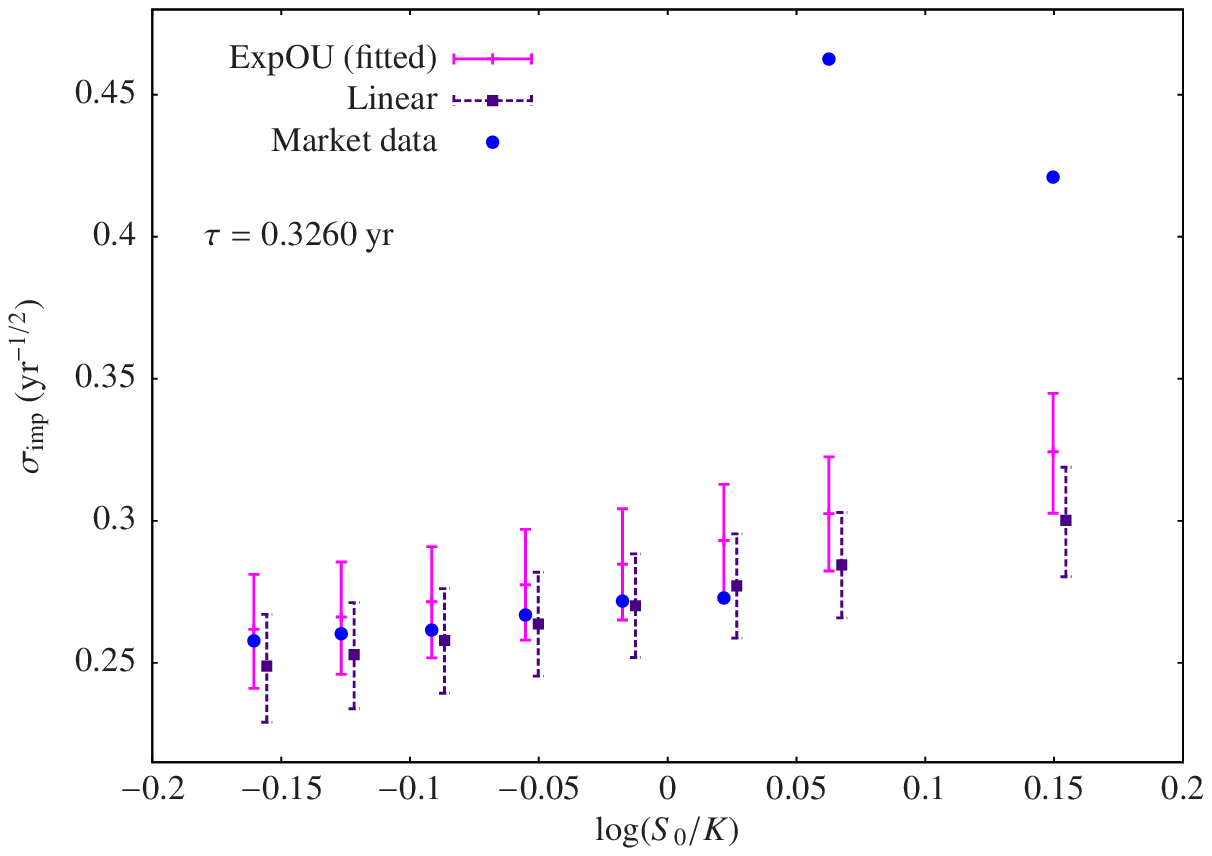}}
   \end{center}
   \caption{Comparison between market implied volatility smiles for Intesa San Paolo and those obtained from the ExpOU and
   Linear models. Parameters values as in the first row of Table 3, as obtained from
   the calibration of the ExpOU model.}
   \label{fig:Csmiles}
\end{figure}

\begin{figure}[h!]
   \begin{center}
      \subfigure
      {\includegraphics[width=0.45\textwidth]{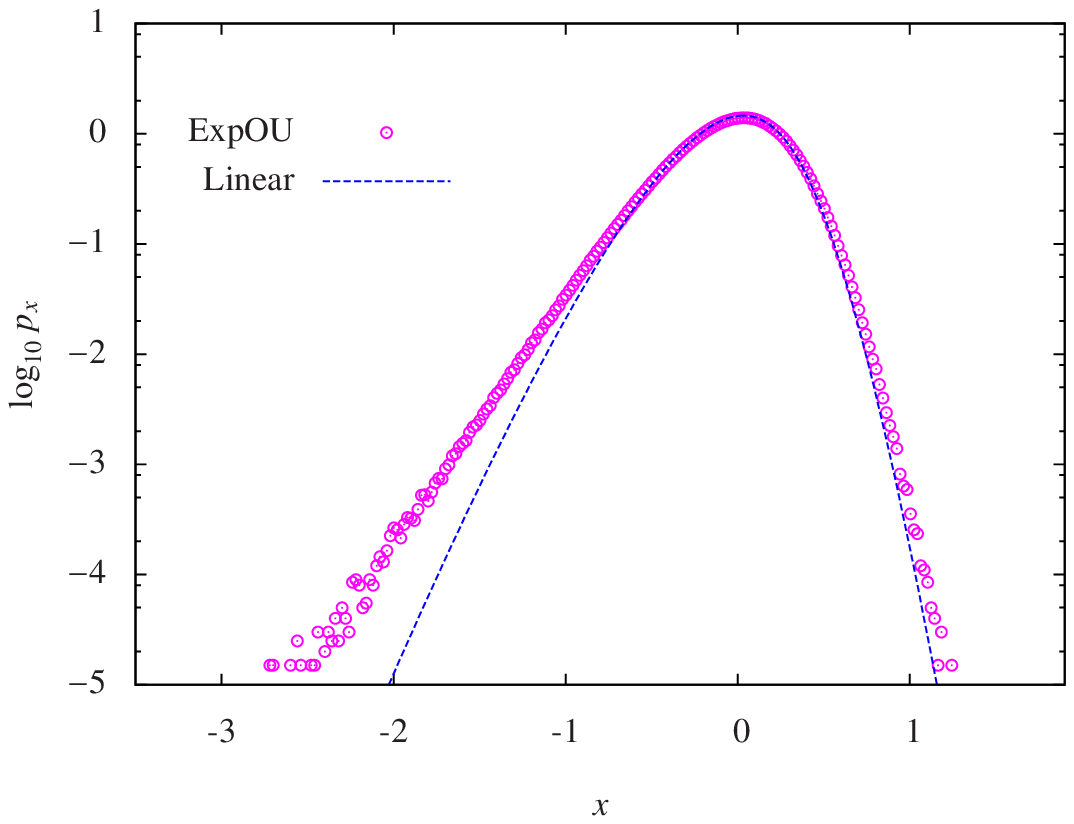}}
      \hspace{5mm}
      {\includegraphics[width=0.45\textwidth]{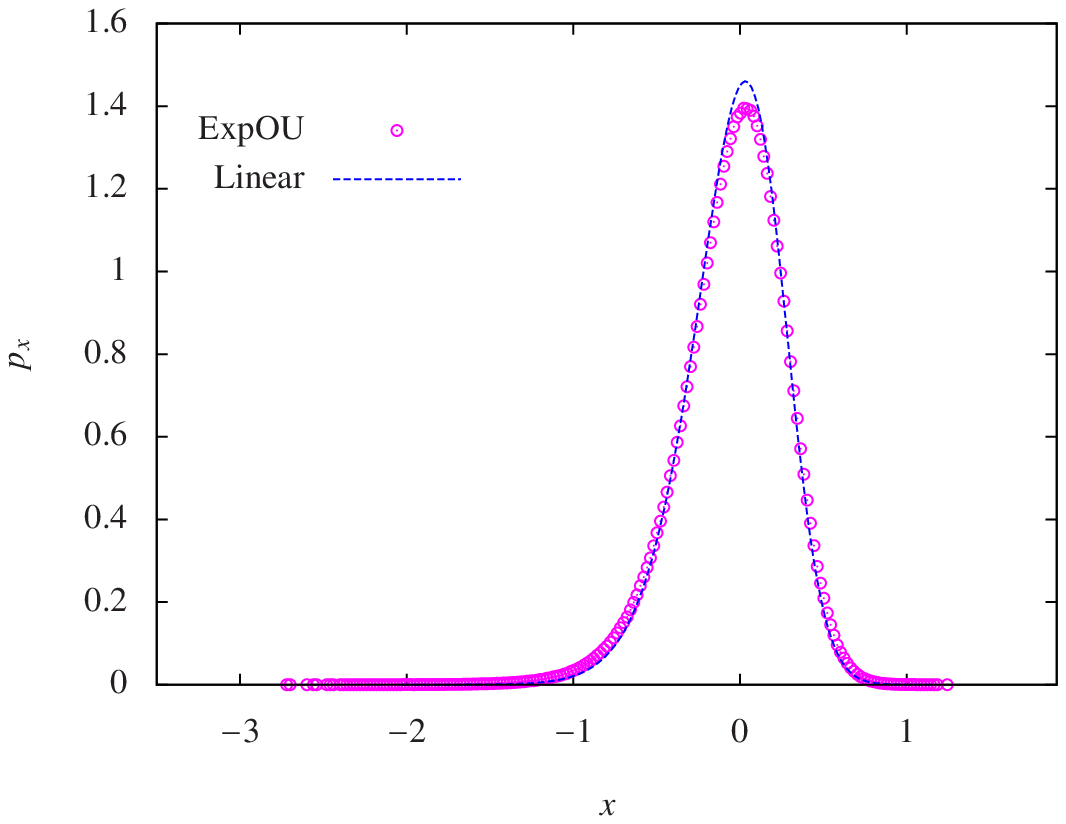}}
   \end{center}
   \caption{Comparison between $p_x(x,\tau|X_0,Z_0)$ with $\tau=1$ for the Linear and ExpOU models in log-linear (left panel) and linear (right panel) scales.
   Parameters values as in the first row of Table 3.}
   \label{fig:PDFcomparison}
\end{figure}

\begin{figure}[h!]
   \begin{center}
      \subfigure
      {\includegraphics[width=0.45\textwidth]{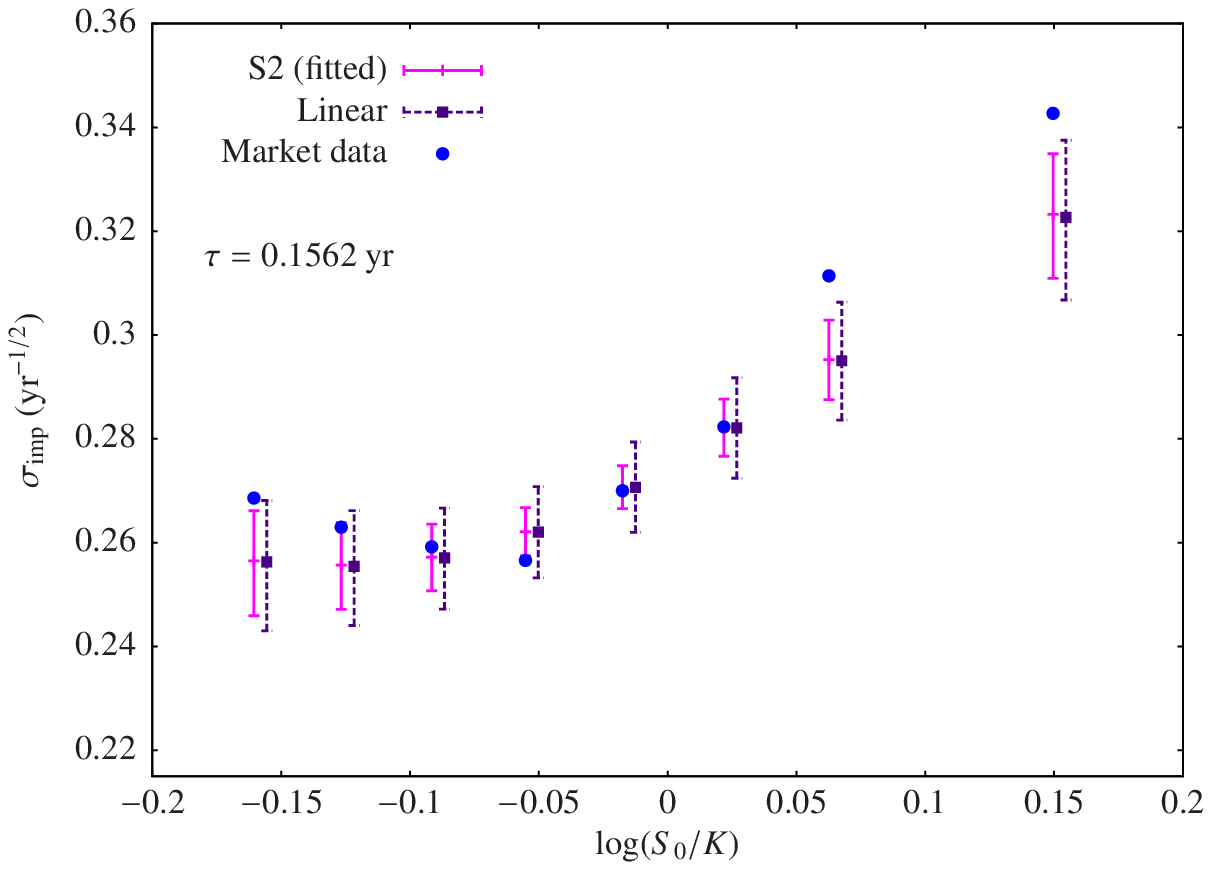}}
      \hspace{5mm}
      {\includegraphics[width=0.45\textwidth]{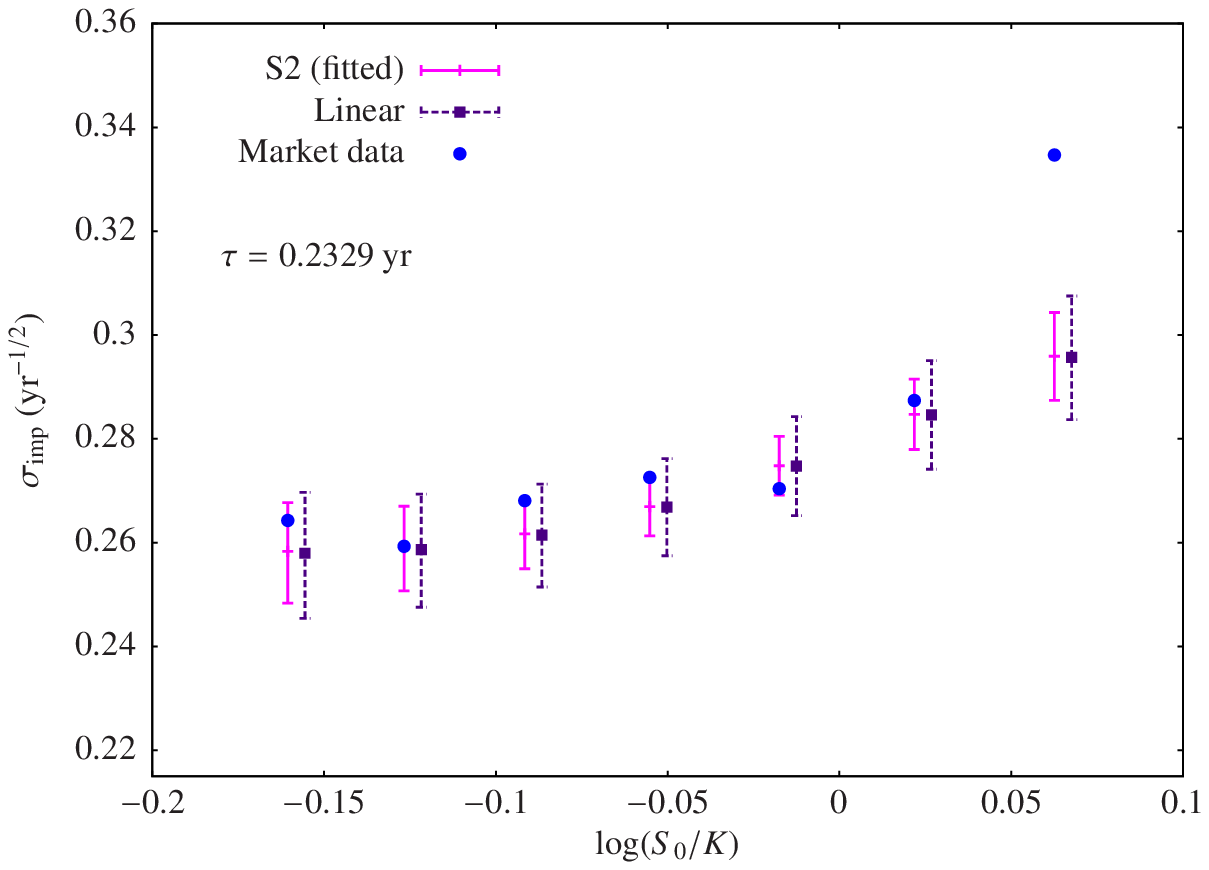}}
      \\
      {\includegraphics[width=0.45\textwidth]{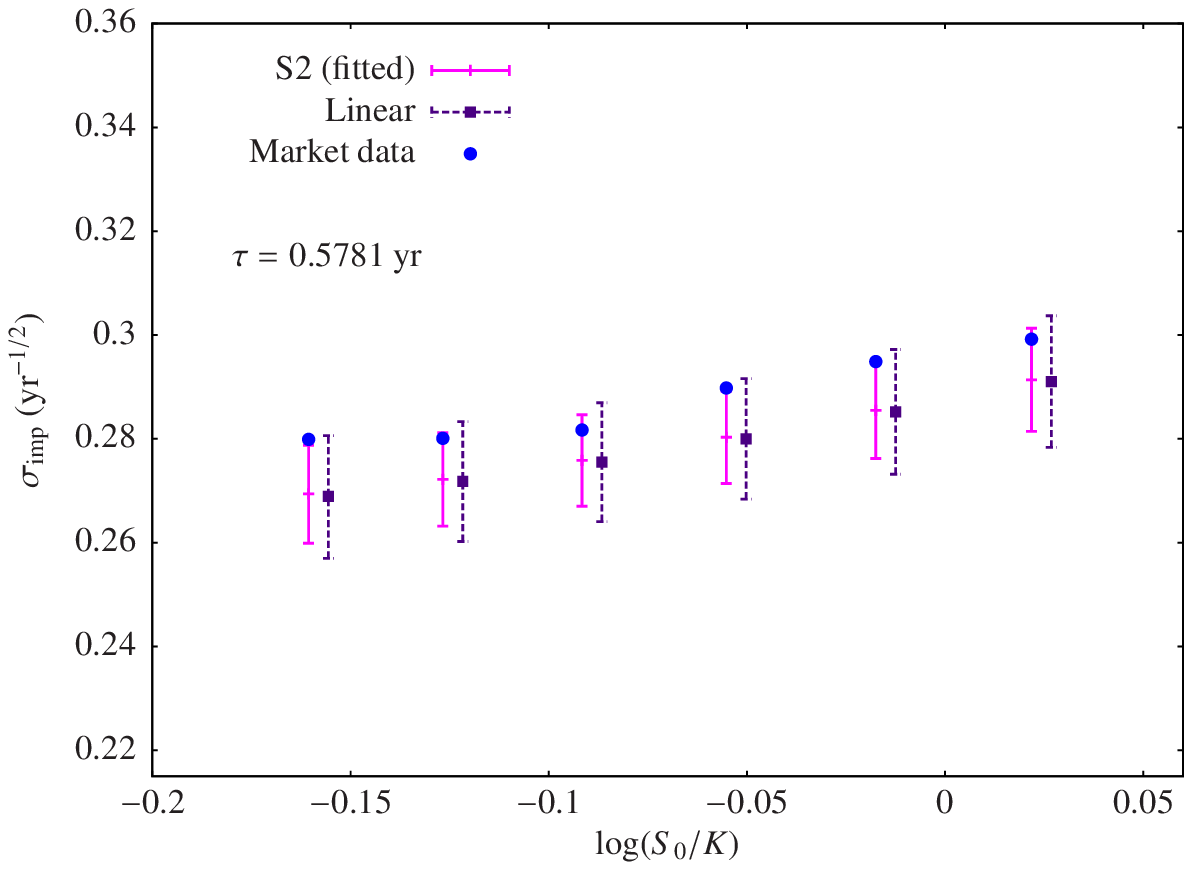}}
      \hspace{5mm}
      {\includegraphics[width=0.45\textwidth]{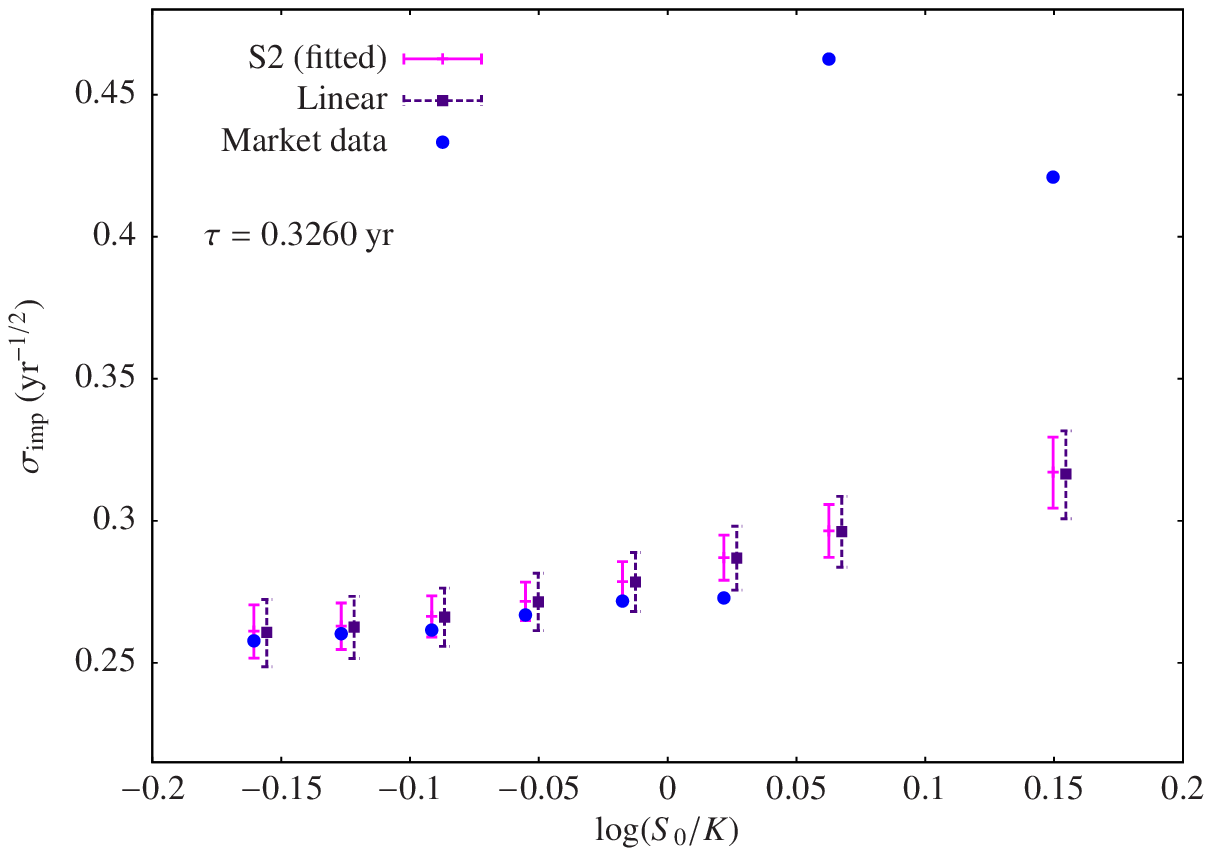}}
   \end{center}
   \caption{Comparison between market implied volatility smiles for Intesa San Paolo and those obtained from the S2 and
   Linear models. Parameters values as in the second row of Table 3, as obtained from
   the calibration of the S2 model.}
   \label{fig:Vsmiles}
\end{figure}

In Fig.~\ref{fig:Csmiles} we present the smiles for the set of parameters corresponding to the ExpOU calibration and we plot both the curves obtained from
MC simulation ($N_{MC}=10^5$) of the exponential model and those computed integrating Eq.\eqref{eq:lewisprice} for the Linear model with the
same parameters values (curves corresponding to the linear case have been shifted rightward). 
In Fig.~\ref{fig:PDFcomparison} we plot the PDF of the Linear model against the one for the ExpOU model computed with trapezoidal integration and MC simulation, 
respectively. Both panels confirm the analysis performed in~\cite{Borm}, showing fatter tails and 
a lower central peak for the histogram of the ExpOU with respect to the PDF of the Linear model.
Even though the value $\beta\sim 13\%$ (see Table~3) is at the edge of the regime allowing the linearisation, 
as far as the volatility smiles obtained from the two models are concerned, we conclude that they are in a good statistical agreement.
In Fig.~\ref{fig:Vsmiles} a comparison analogous to the one in Fig.~\ref{fig:Csmiles} for the S2 parameters is reported, 
revealing again the statistical agreement. With respect
to Fig.~\ref{fig:Csmiles}, the narrower error bars reflect the fact that parameters fitting has been performed exploiting the available
analytical information. Actually, the MC simulation involved both in the calibration and the price computation for ExpOU introduces
an additional statistical uncertainty.
\section{Conclusions and Perspectives}\label{sec:Concl}

This paper deals with the problem of option pricing under stochastic volatility and calibration
to market smiles.
The main focus is on a specific SVM where the dynamics of financial log-returns is driven by linear drift and diffusion coefficients,
which we show to be the limit case of the exponential Ornstein-Uhlenbeck and the Stein-Stein models
under the low volatility fluctuation regime. 
The analytical contribution we provide is the exact characterization of the characteristic function of the Linear model
under risk neutrality. Following the market practice, we calibrate the model on a sample of implied volatilities by means of
the two step calibration procedure detailed in Section~\ref{sec:NumRes}. In this regard, the knowledge of the analytical expressions
of the cumulants substantially reduces the computational effort required to fix the parameters values. For the considered data, we
are able to quantitatively asses the capability of the model to reproduce the market volatility smiles,
finding a statistically significant agreement with the empirical curves.
By means of the same procedure, we also compute the implied volatility values under the ExpOU and S2 models after calibration to evaluate
the accuracy of the linear approximation. In both cases, the propagation of the statistical uncertainty results in an error band which
clearly reveals that the three models are all in full agreement. In particular, the measured value of $\beta$ for the ExpOU model justifies
its linearisation, whose degree of analytical tractability makes it more desirable in view of option pricing.

In this work, we exploited the Greeks to evaluate the statistical uncertainty of the reconstructed volatility smiles;
as a future perspective, it would be interesting to analyze the sensitivity with respect to movements of
market volatility curves, which is what is done in financial practice by traders to hedge option positions.
We also aim at applying the Linear model considered here in the context of market risk measures, like Value-at-Risk and Expected Shortfall,
exploiting the knowledge of the analytical CF following the guidelines traced by~\cite{Bormetti_etal:2010}. For risk management purposes, it would 
be interesting to analytically characterize the PDF tails. Indeed, the numerical convergence of the integral in Eq.~\eqref{eq:lewisprice} excludes a power law decay,
but an explicit analytical result such as for the Heston~\cite{Dragulescu:2002} and the ExpOU models~\cite{Jourdain:2004} 
discerning between exponential, Gaussian or different scalings is still lacking. 

\section*{Acknowledgments}

We wish to acknowledge the anonymous referee for fruitful comments and for giving us the opportunity, through
his criticism, of improving the paper in several respects. We also thank Enrico Melchioni, Guido Montagna, Oreste Nicrosini, Andrea Pallavicini and
Fulvio Piccinini for their suggestions. We are grateful to FMR Consulting for having provided the market data.


\appendix \section{Linear model cumulants} \label{AppA}

In the following we report the analytical expressions of the cumulants of the Linear model.
\begin{equation*}
	k_{1,\tau} = -\frac{m^2}{2}\int_0^\tau\mathcal{M}(\tau')\ud\tau'+\frac{m^2}{\alpha}(Z_0-1) (e^{-\alpha \tau} -1)-\frac{m^2}{2\alpha}\alpha \tau + X_0,
\end{equation*}
\begin{align*}
  k_{2,\tau} = \frac{1}{4} \frac{m^2}{\alpha} &\bigg\{
  -2\left( \frac{km}{\alpha} \right)^2 \left[ e^{-2\alpha \tau}-4e^{-\alpha \tau}-2\alpha \tau+3 \right]
  +\frac{k^2}{\alpha} \left[ e^{-2\alpha \tau}+2\alpha \tau-1 \right]\nonumber\\
  &-2(Z_0-1)^2\left[ e^{-2\alpha \tau}-1 \right]-8(Z_0-1)\left[ e^{-\alpha \tau}-1 \right]
  +4\alpha \tau \bigg\}\nonumber\\
  +2~\frac{km^3}{\alpha^2}\rho~ &\bigg\{(Z_0-1)\left[ e^{-\alpha \tau}+\alpha \tau e^{-\alpha \tau}-1 \right]-
  \left[ e^{-\alpha \tau}+\alpha \tau-1 \right] \bigg\},
\end{align*}
\begin{align*}
  k_{3,\tau} = \frac{3}{2} \frac{k^2m^3}{\alpha^3} &\bigg\{ (Z_0-1)\left[ e^{-3\alpha \tau}-2e^{-2\alpha \tau}
  +e^{-\alpha \tau}(3+2\alpha \tau)-2\right]+2\left[ e^{-2\alpha \tau}-4e^{-\alpha \tau}-2\alpha \tau+3
  \right] \bigg\}\nonumber\\
  +\frac{3}{2}\frac{km^3}{\alpha^2}\rho~ &\bigg\{ \left( \frac{km}{\alpha} \right)^2
  \left[ -e^{-2\alpha \tau}(3+2\alpha \tau)+4e^{-\alpha \tau}(3+\alpha \tau)+4\alpha \tau-9 \right]
  +\frac{k^2}{\alpha}\left[ e^{-2\alpha \tau}(1+\alpha \tau)+\alpha \tau-1 \right]\nonumber\\
  -&(Z_0-1)^2\left[ e^{-2\alpha \tau}(1+2\alpha \tau)-1 \right]+2(Z_0-1)
  \left[ e^{-2\alpha \tau}-2e^{-\alpha \tau}(2+\alpha \tau)+3 \right]+
  4\left[ e^{-\alpha \tau}+\alpha \tau-1 \right]\bigg\}\nonumber\\
  +3~ \frac{k^2m^4}{\alpha^3}\rho^2 &\bigg\{ (Z_0-1)\left[ e^{-\alpha \tau}(2+2\alpha \tau+\alpha^2 \tau^2)
  -2\right]-2\left[ e^{-\alpha \tau}(2+\alpha \tau)+\alpha \tau-2 \right] \bigg\},
\end{align*}
\begin{align*}
  k_{4,\tau} = 3~\frac{k^2m^4}{\alpha^3}&\bigg\{ 
  \frac{1}{2}\left( \frac{km}{\alpha} \right)^2
  \left[ -e^{-4\alpha \tau}+4e^{-3\alpha \tau}-4e^{-2\alpha \tau}(3+\alpha \tau)+4e^{-\alpha \tau}(7+2\alpha \tau)
  +8\alpha \tau-19\right]\nonumber\\
  &+\frac{k^2}{8\alpha}\left[ e^{-4\alpha \tau}+4e^{-2\alpha \tau}(1+2\alpha \tau)+4\alpha \tau-5 \right]
  -\frac{1}{2}(Z_0-1)^2\left[ e^{-4\alpha \tau}+4\alpha \tau e^{-2\alpha \tau}-1 \right]\nonumber\\
  &+2(Z_0-1)\left[ -e^{-3\alpha \tau}+2e^{-2\alpha \tau}-e^{-\alpha \tau}(3+2\alpha \tau)+2 \right]+
  2\left[ -e^{-2\alpha \tau}+4e^{-\alpha \tau}+2\alpha \tau-3 \right]\bigg\}\nonumber\\
  +6~ \frac{k^3m^5}{\alpha^4}\rho ~&\bigg\{ (Z_0-1)\left[ 3e^{-3\alpha \tau}(1+\alpha \tau)-
  2e^{-2\alpha \tau}(3+2\alpha \tau)+e^{-\alpha \tau}(9+7\alpha \tau+2\alpha^2 \tau^2)-6\right]\nonumber\\
  & +\left[ -e^{-3\alpha \tau}+2e^{-2\alpha \tau}(5+2\alpha \tau)-e^{-\alpha \tau}(35+10\alpha \tau)-12\alpha \tau+26
  \right]\bigg\}\nonumber\\
  +3~ \frac{k^2m^4}{\alpha^3}\rho^2 &\bigg\{ 4\left( \frac{km}{\alpha} \right)^2
  \left[ -e^{-2\alpha \tau}(3+3\alpha \tau+\alpha^2 \tau^2)+e^{-\alpha \tau}(12+6\alpha \tau+\alpha^2 \tau^2)+3\alpha \tau-9 \right]\nonumber\\
  & +\frac{k^2}{\alpha}\left[ e^{-2\alpha \tau}(3+4\alpha \tau+2\alpha^2 \tau^2)+2\alpha \tau-3 \right]
  -2(Z_0-1)^2\left[ e^{-2\alpha \tau}(1+2\alpha \tau+2\alpha^2 \tau^2)-1 \right]\nonumber\\
  & +4(Z_0-1)\left[ 2e^{-2\alpha \tau}(1+\alpha \tau)-e^{-\alpha \tau}(6+4\alpha \tau+\alpha^2 \tau^2)+4 \right]\nonumber\\
  & -2\left[ e^{-2\alpha \tau}-4e^{-\alpha \tau}(3+\alpha \tau)-6\alpha \tau+11 \right]\bigg\}\nonumber\\
  4~ \frac{k^3m^5}{\alpha^4}\rho^3 &\bigg\{ (Z_0-1)\left[ e^{-\alpha \tau}(6+6\alpha \tau+3\alpha^2 \tau^2+\alpha^3 \tau^3)
  -6\right]-3\left[ e^{-\alpha \tau}(6+4\alpha \tau+\alpha^2 \tau^2)+2\alpha \tau-6 \right] \bigg\} ~ .
\end{align*}
From the asymptotic expansions $k_{2,\tau} \sim -m^2(2Z_0-1)\tau/2$,
$k_{3,\tau} \sim 3km^3Z_0^2\rho \tau^2$, and $k_{4,\tau}\sim 4k^2m^4(1+2\rho^2)Z_0^2\tau^3$ 
when $\tau\rightarrow 0^+$, we can infer the leading behaviour of skewness and kurtosis at the origin
\begin{equation*}
  \zeta_{\tau} \sim 3\frac{k\rho}{Z_0}\sqrt{\tau} \quad \text{and} \quad \kappa_{\tau}\sim 4\frac{k^2(1+2\rho^2)}{Z_0^2}\tau.
\end{equation*}


\begin{thebibliography}{00}
   \bibitem{BS:1973} F. Black and M. Scholes, The pricing of options and corporate liabilities, \emph{J. Polit. Economy} {\bf 81} (1973) 637.
   \bibitem{Merton:1973} R. Merton, Theory of rational option pricing, \emph{Bell J. Econ. Management Sci.} {\bf 4} (1973) 141.
   \bibitem{Heston:1993} S. Heston, A closed-form solution for options with stochastic volatility with applications to bond and currency options,
      \emph{Rev. Finan. Stud.} {\bf 6} (1993) 327.
   \bibitem{Stein_Stein:1991} E.M. Stein and J.C. Stein, Stock price distributions with stochastic volatility: An analytic approach, 
      {\it Rev. Finan. Stud.} {\bf 4} (1991) 727.
   \bibitem{SZ} R. Sch\"obel and J. Zhu, Stochastic volatility with an Ornstein Uhlenbeck process: An extension, \emph{Europ. Finance Rev.} {\bf 4}
      (1999) 23.
   \bibitem{Hull:1987} J. Hull and A. White, The pricing of options on asset with stochastic volatilities, \emph{J. Finance}
      {\bf 42} (1987) 281.
   \bibitem{Scott:1987} L. Scott, Option pricing when the variance changes randomly: Theory, estimators and applications, \emph{J. Finan. Quant. Anal.} {\bf 22} (1987) 419.
   \bibitem{Micciche:2002} S. Miccich\`e, G. Bonanno, F. Lillo and R. N. Mantegna, Volatility in financial markets: Stochastic models and empirical results.
      \emph{Physica A} {\bf 314} (2002) 756.
   \bibitem{Lipton:2008} A. Lipton and A. Seppt, Stochastic volatility models and Kelvin waves, \emph{J. Phys. A} {\bf 41} (2008) 344012.
   \bibitem{Mitra:2009} S. Mitra, A review of volatility and option pricing. \emph{Preprint} available at: http://arxiv.org/pdf/0904.1292.
   \bibitem{MP:2006} J. Masoliver and J. Perell\'o, Multiple time scales and the exponential Ornstein-Uhlenbeck stochastic volatility
      model, \emph{Quant. Finance} {\bf 6} (2006) 423.
   \bibitem{Borm} G. Bormetti, V. Cazzola, G. Montagna and O. Nicrosini, The probability distribution of returns in the exponential
      Ornstein - Uhlenbeck model, {\it J. Stat. Mech.} (2008) P11013.
   \bibitem{Carr:1999} P. Carr and D.B. Madan, Option valuation using the Fast Fourier Transform, \emph{J. Comp. Finance} {\bf 2} (1999) 61.
   \bibitem{Lewis:2001} A. L. Lewis, A simple option formula for general jump - diffusion and other exponential L\'evy processes, 
      Envision Financial Systems and OptionCity.net Technical Report (2001). Available at http://www.optioncity.net.
   \bibitem{Lipton:2001} A. Lipton, Mathematical Methods For Foreign Exchange: A Financial Engineer's Approach, World Scientific Publishing (2001).
   \bibitem{Lord:2008} R. Lord and C. Kahl, Complex logarithms in Heston-like models. Available at http://ssrn.com/abstract=1105998.
   \bibitem{Perello:2007} J. Perell\'o, Market memory and fat tail consequences in option pricing on the
      expOU stochastic volatility, \emph{Physica A} {\bf 382} (2007) 213.
   \bibitem{PSM} J. Perell\'o, R. Sircar and J. Masoliver, Option pricing under stochastic volatility: The exponential Ornstein-Uhlenbeck model,
      {\it J. Stat. Mech.} (2008) P06010.
   \bibitem{FPS} J. P. Fouque, G. Papanicolau and K. R. Sircar,	Derivatives in financial markets with stochastic volatility, Cambridge University Press (2006).
   \bibitem{Masoliver_Perello:2002} J. Masoliver and J. Perell\'o, A correlated stochastic volatility model measuring leverage and other stylized facts,
      {\it Int. J. Theoretical Appl. Finance} {\bf 5} (2002) 541.
   \bibitem{Veltman:1979} G. 't Hooft and M. Veltman, Scalar one-loop integrals, \emph{Nucl. Phys. B} \textbf{153} (1979) 365.
   \bibitem{Mikhailov:2004} S. Mikhailov and U. N\"ogel, Heston's stochastic volatility model: Implementation, calibration and some extensions, in  
      {\it The Best of Wilmott 1: Incorporating the Quantitative Finance Review}, P. Wilmott (ed.), John Wiley and Sons (2004).
   \bibitem{Galluccio:2005} S. Galluccio and Y. Le Cam, Implied calibration of stochastic volatility jump diffusion
      models, (2005). Available at http://129.3.20.41/eps/fin/papers/0510/0510028.pdf.
   \bibitem{Forde:2009} M. Forde, A. Jaquier and A. Mijatovic, Asymptotic formulae for implied volatility in the Heston model, \emph{arXiv0911.2992} (2009). 
   \bibitem{Backus:2004} D. K. Backus, S. Foresi and L. Wu, Accounting for biases in Black-Scholes, {\it Technical report} (2004). 
      Available at SSRN: http://ssrn.com/abstract=585623.
   \bibitem{Bouchaud:2000} J. P. Bouchaud and M. Potters, Theory of Financial Risk and Derivative Pricing: From Statistical Physics to Risk Management, 
      Cambridge University Press (2000).
   \bibitem{Ciliberti:2008} S. Ciliberti, J. P. Bouchaud and M. Potters, Smile dynamics: A theory of the implied leverage effect. {\it Preprint} 
      available at http://lanl.arxiv.org/abs/0809.3375.
   \bibitem{Press:1989} W. H. Press, B. P. Flannery, S. A. Teukolsky and W. T. Vetterling, Numerical recipes -- The art of scientific computing, Cambridge 
      University Press, New York (1989).
   \bibitem{Minuit} F. James, MINUIT Function Minimization and Error Analysis Reference Manual, CERN Geneva, Switzerland (1998). 
      Available at: http://wwwasdoc.web.cern.ch/wwwasdoc/minuit/minmain.html.
   \bibitem{Bormetti_etal:2010} G. Bormetti, V. Cazzola, G. Livan, G. Montagna and O. Nicrosini, A generalized Fourier transform approach to risk measures, 
	   {\it J. Stat. Mech.} (2010) P01005.   
   \bibitem{Dragulescu:2002} A. A. Drag\u{u}lescu and V. M. Yakovenko, Probability distribution of returns in the Heston model with stochastic volatility, 
      {\it Quant. Finance} {\bf 2} (2002) 443.
   \bibitem{Jourdain:2004} B. Jourdain, Loss of martingality in asset price models with lognormal stochastic volatility, {\it Preprint CERMICS 2004-267}
	   available at http://cermics.enpc.fr/~jourdain/publications.html.
\end{thebibliography}
\end{document}